\documentclass[12pt,revtex4-1]{article}
\usepackage{graphicx}
\usepackage{epsfig,amsmath,amssymb} 
\usepackage{dcolumn}
\usepackage{bm}
\usepackage{subfigure}
\usepackage{float}
\usepackage{color}
\usepackage{longtable}
\usepackage{cite}

\newcommand{\be}{\begin{equation}}
\newcommand{\ee}{\end{equation}}

\newcommand{\ra}{\rangle}
\newcommand{\la}{\langle}
\newcommand{\bit}{\begin{itemize}}
\newcommand{\eit}{\end{itemize}}
\newcommand{\bea}{\begin{eqnarray}}
\newcommand{\eea}{\end{eqnarray}}

\usepackage{epsfig}

\begin{document}
\title
{The frustrated spin-$1/2$ $J_1$-$J_2$ Heisenberg ferromagnet on the square lattice: 
Exact diagonalization and Coupled-Cluster study
}
\author
{J. Richter$^{1}$, R. Darradi$^{2}$, J.~Schulenburg$^3$, 
D.J.J. Farnell$^{4}$, and H. Rosner$^{5}$  \\
\small{$^{1}$Institut f\"ur Theoretische Physik, Universit\"at
Magdeburg, D-39016 Magdeburg, Germany }\\
\small{$^{2}$ Institut f\"ur Theoretische Physik, Technische Universit\"at
Braunschweig, D-38106 Braunschweig, Germany}\\
\small{$^{3}$Universit\"{a}tsrechenzentrum,
             Universit\"{a}t Magdeburg, D-39016 Magdeburg, Germany}\\
\small{$^{4}$Academic Department of Radiation Oncology,
University of Manchester, United Kingdom}\\
\small{$^{5}$Max-Planck Institut f\"{u}r Chemische Physik fester
Stoffe, D-01187 Dresden, Germany}
}

\date{\today}
       
\maketitle

\begin{abstract}
We investigate  the ground-state magnetic order  of the spin-$1/2$ $J_1$-$J_2$
Heisenberg model  on the square lattice with ferromagnetic nearest-neighbor
exchange $J_1<0$ and frustrating  
antiferromagnetic  next-nearest neighbor exchange $J_2>0$.
We use the coupled-cluster method to high orders of approximation and Lanczos
exact diagonalization of finite lattices  of up to $N=40$ sites in order to 
calculate the ground-state energy, the spin-spin correlation functions, 
and the magnetic order parameter. 
We find that the transition point at which the ferromagnetic ground
state disappears is given by $J_2^{c1}=0.393|J_1|$ (exact diagonalization) and
$J_2^{c1}=0.394|J_1|$ (coupled-cluster method).     
We compare our results for ferromagnetic $J_1$ with established results for 
the spin-$1/2$ $J_1$-$J_2$ Heisenberg model with antiferromagnetic 
$J_1$. We find that both models (i.e., ferro- and antiferromagnetic $J_1$) 
behave similarly for large $J_2$, although significant differences between them
are observed for $J_2/|J_1| \lesssim 0.6$.
Although the semiclassical collinear magnetic long-range order breaks down at
$J_2^{c2}\approx 0.6J_1$ for antiferromagnetic $J_1$, we do not find a similar
breakdown of this kind of long-range order until $J_2 \sim 0.4|J_1|$ for the 
model with ferromagnetic $J_1$.
Unlike the case for antiferromagnetic $J_1$, if an intermediate disordered phase 
does occur between the phases exhibiting semiclassical collinear stripe order 
and ferromagnetic order for ferromagnetic $J_1$ then it is likely to be over a very 
small range below $J_2 \sim 0.4|J_1|$.
\end{abstract}

PACS codes:\\
75.10.Jm Quantized spin models\\
75.45.+j Macroscopic quantum phenomena in magnetic systems\\
75.10.Kt Quantum spin liquids, valence bond phases and related phenomena\\

\section{Introduction}

The spin-$1/2$ Heisenberg antiferromagnet with nearest-neighbor (NN)
$J_1$ and frustrating next-nearest-neighbor (NNN) $J_2$ coupling
($J_1$-$J_2$ model) on the square lattice has attracted much attention
during the last twenty years (see, e.g.,
Refs.~\cite{chandra88,dagotto89,figu90,schulz,richter93,richter94,Trumper97,bishop98,capriotti01,siu01,singh03,Sir:2006,Schm:2006,darradi08,bishop08,xxz,singh2009,cirac2009,ED40,fprg}
and references therein). This model may serve as a canonical model to
study the interplay of frustration effects and quantum fluctuations as
well as quantum phase transitions driven by frustration.

The corresponding Hamiltonian reads
\begin{equation}
H = J_{1}\sum_{\langle i,j
\rangle}{\bf s}_i \cdot {\bf s}_j
  +  J_{2}\sum_{\langle\langle i,j
\rangle\rangle}{\bf s}_i \cdot {\bf s}_j,
\end{equation}
where the sums over $\langle i,j \rangle$ and $\langle\langle i,j
\rangle\rangle$ run over all NN and NNN pairs, respectively, counting
each bond once.

Recently, several quasi-two-dimensional magnetic materials with a
ferromagnetic (FM) NN coupling $J_1<0$ and an antiferromagnetic (AFM)
NNN coupling $J_2>0$ have been studied experimentally, e.g.,
Pb$_2$VO(PO$_4$)$_2$ \cite{kaul04,jmmm07,carretta2009,enderle},
(CuCl)LaNb$_2$O$_7$ \cite{kageyama05}, SrZnVO(PO$_4$)$_2$
\cite{rosner09,rosner09a,enderle}, and BaCdVO(PO$_4$)$_2$
\cite{nath2008,carretta2009,rosner09}.
Due to a quite large AFM coupling $J_2$ these materials are driven out
of the FM phase.  These experimental studies have
stimulated a series of theoretical investigations of the ground state
(GS) and thermodynamic properties of the FM $J_1$-$J_2$ model,
i.e. the model with FM NN exchange $J_1$ and frustrating AFM NNN
exchange $J_2$
\cite{shannon04,shannon06,sindz07,schmidt07,schmidt07_2,sousa,shannon09,sindz09,haertel10}.
It was found that the FM GS for the spin-$1/2$ model breaks
down at $J_2 = J^{c1}_2 \approx 0.4|J_1|$
\cite{shannon04,shannon06,sindz07,sousa,haertel10,dmitriev97}. 
Note that for the classical model ($s \to \infty$) the corresponding
transition point is at $J_2=0.5|J_1|$.  Moreover, for sufficiently
large $J_2 > J^{c2}_2$ a magnetically long-range ordered collinear stripe GS
appears.  According to
Refs. \cite{shannon06,sindz07,shannon09,sindz09} this second critical
frustration strength is $J^{c2}_2 \approx 0.6|J_1|$, and both
magnetically ordered phases are separated by a
magnetically disordered phase with enhanced nematic correlations.
Note that this collinear stripe GS phase, as well as the values of
$J^{c1}_2$ and $J^{c2}_2$ reported in the literature, are similar to
the observation for the corresponding AFM model, i.e. the model with
AFM $J_1$
\cite{chandra88,schulz,richter94,Trumper97,capriotti01,singh03,
Sir:2006,Schm:2006,darradi08,xxz,singh2009,cirac2009,ED40,fprg}.

As we know from the investigation of the AFM $J_1$-$J_2$ model the
study of the GS phases in the strong frustration regime, i.e.  $J_2
\sim 0.5 J_1$, is one of the hardest problems in the field of
frustrated quantum magnetism. In the meantime, scores of papers have
dealt with this problem. On the other hand, for the FM $J_1$-$J_2$
model so far only a few studies exist. Hence, further investigations
using alternative approaches are highly desirable to confirm or to
query the existing results.

Due to frustration, highly efficient quantum Monte-Carlo codes, such
as the stochastic series expansion suffer from the minus sign
problem. Therefore, many other approximate methods, e.g.\ the Green's
function method \cite{siu01,Schm:2006,haertel10}, the series expansion
\cite{singh03,Sir:2006,singh2009}, the Schwinger boson approach
\cite{Trumper97}, variational techniques \cite{capriotti01}, the functional
renormalization group technique \cite{fprg} as well
as the projected entangled pair states method \cite{cirac2009} were
used to study the GS phases of the model with AFM $J_1$.

In the present paper we calculate the GS energy,  spin-spin correlation
functions, and the magnetic order
parameter of the collinear stripe order using two completely different
methods, namely a large-scale exact diagonalization (ED) of finite
lattices up to $N=40$ sites (see Sec.~\ref{ED}) and the
coupled-cluster method (CCM) (see Sec.~\ref{ccm}). Both methods have
been successfully applied to the AFM $J_1$-$J_2$ model, see
e.g. Refs.~\cite{dagotto89,figu90,schulz,richter93,richter94,ED40} for
the ED method and
Refs.~\cite{bishop98,Schm:2006,darradi08,bishop08,xxz} for the CCM.

Henceforth we set $|J_1|=1$ if not stated otherwise explicitly.

\section{Results of Lanczos exact diagonalization for finite square lattices} 
\label{ED}

The paper of Schulz and co-workers\cite{schulz}, presenting 
results for finite lattices up to $N=36$ sites for the
first time has set a
benchmark for ED studies of quantum Heisenberg magnets. Frequently,
such numerical exact results are also used to test new approximate
methods, see, e.g. Refs.~\cite{Kash:2001,Mune:2007,darradi08}.  Due to
the progress in the computer hardware and the increased efficiency in
programming, very recently the GS and low-lying excitations of the
unfrustrated (i.e., $J_2=0$)\cite{wir04} as well as the frustrated
spin-$1/2$ HAFM \cite{ED40} and of a spin-$1/2$ Heisenberg model with
ring exchange \cite{Lauch:2004} have been calculated by the Lanczos
algorithm for a square lattice with $N=40$ sites.  The largest
two-dimensional quantum spin model for which the GS has been
calculated so far is the spin-$1/2$ HAFM on the star lattice with
$N=42$ sites \cite{star04}.  Note, however, that for sectors of $S_z >
0$ which are relevant for finite magnetic fields much larger system
sizes can be treated by ED, see, e.g., Ref.~\cite{prl02}.

\begin{table}
\begin{center}
\caption
{Singlet GS energy per site $E_0/N$ and square of order parameter
$M_N^2(0,\pi)$, cf. Eq.~(\ref{M_squared}),
for the $J_1$-$J_2$ model with $J_1=-1$ on
finite square lattices of $N=32$, $36$ and $40$ sites.
\label{tab1}}
\vspace{5mm}
\begin{tabular}[t]{|c|c|c|c|c|c|c|c|c|c|c|c|c|c|c|c|c|} \hline
       & $N=32$ & $N=32$ & $N=36$ &  $N=36$& $N=40$ &  $N=40$ \\ \hline \hline
 $J_2$ & $E_0/N$ & $M_N^2({0,\pi})$ & $E_0/N$ & $M_N^2({0,\pi})$ & $E_0/N$ & $M_N^2({0,\pi})$ \\ \hline \hline
 0.40  &   -0.31284253  &  0.11167341  &  -0.31431081 &   0.08925821  & -0.31127637   &   0.10687524    \\  
 0.45  &   -0.34234578  &  0.12363200  &  -0.34259033 &   0.11439947  & -0.34062128   &   0.11332095    \\  
 0.50  &   -0.37406800  &  0.12617129  &  -0.37369533 &   0.11854042  & -0.37174835   &   0.11581048    \\  
 0.55  &   -0.40673497  &  0.12730259  &  -0.40577614 &   0.12049011  & -0.40374930   &   0.11719238    \\ 
 0.60  &   -0.43994975  &  0.12793318  &  -0.43841728 &   0.12170653  & -0.43628038   &   0.11810286    \\ 
 0.65  &   -0.47352394  &  0.12833506  &  -0.47143169 &   0.12256200  & -0.46917022   &   0.11876286    \\ 
 0.70  &   -0.50735237  &  0.12861459  &  -0.50471467 &   0.12320526  & -0.50231993   &   0.11927048    \\ 
 0.80  &   -0.57553353  &  0.12898259  &  -0.57184472 &   0.12411474  & -0.56916790   &   0.12000667    \\ 
 0.90  &   -0.64418634  &  0.12921694  &  -0.63949153 &   0.12472863  & -0.63652042   &   0.12051714    \\ 
 1.00  &   -0.71315309  &  0.12937976  &  -0.70748822 &   0.12516916  & -0.70421577   &   0.12089143    \\
 \hline 
\end{tabular}
\end{center}
\end{table}

\begin{table}
\begin{center}
\caption
{Spin-spin correlation functions 
$\langle {\bf s}_0\cdot{\bf s}_{\bf R}\rangle$ for the $J_1$-$J_2$ with
$J_1=-1$ model on
the finite square lattice of $N=40$ sites.
\label{tab2}}
\vspace{5mm}
\begin{tabular}[t]{|c|c|c|c|c|c|c|c|c|c|c|c|c|c|c|c|c|} \hline
        & $\langle {\bf s}_0 \cdot{\bf s}_{\bf R}\rangle$ & $\langle {\bf s}_0 \cdot{\bf s}_{\bf R}\rangle$ & $\langle {\bf s}_0 \cdot{\bf s}_{\bf R}\rangle$ & $\langle {\bf s}_0 \cdot{\bf s}_{\bf R}\rangle$ & $\langle {\bf s}_0 \cdot{\bf s}_{\bf R}\rangle$ & $\langle {\bf s}_0 \cdot{\bf s}_{\bf R}\rangle$  \\ \hline 
 $J_2$ &  ${\bf R}=(1,0)$ & ${\bf R}=(2,0)$ & ${\bf R}=(3,0)$ & ${\bf R}=(3,1)$ & ${\bf R}=(2,1)$ & ${\bf R}=(1,1)$  \\ \hline \hline
 0.40    &  0.044832   & 0.183933   & 0.012219   & -0.194442  & -0.040218  & -0.277017   \\  
 0.45    &   0.033144  &  0.196362  &  0.014787  & -0.199197  & -0.033852  & -0.304812   \\  
 0.50    &   0.027648  &  0.201318  &  0.014973  & -0.200208  & -0.029595  & -0.316452   \\  
 0.55    &   0.024204  &  0.204078  &  0.014523  & -0.200667  & -0.026460  & -0.323037   \\ 
 0.60    &   0.021753  &  0.205887  &  0.013884  & -0.201012  & -0.024012  & -0.327312   \\ 
 0.65    &   0.019875  &  0.207189  &  0.013203  & -0.201327  & -0.022023  & -0.330324   \\ 
 0.70    &   0.018363  &  0.208185  &  0.012537  & -0.201621  & -0.020367  & -0.332568   \\ 
 0.80    &   0.016044  &  0.209616  &  0.011328  & -0.202149  & -0.017745  & -0.335676   \\ 
 0.90    &   0.014313  &  0.210600  &  0.010293  & -0.202593  & -0.015750  & -0.337719   \\ 
 1.00    &   0.012954  &  0.211320  &  0.009417  & -0.202962  & -0.014172  & -0.339156   \\
 \hline 
\end{tabular}

\vspace{0.1cm}
\begin{tabular}[t]{|c||c|c|c|c|c|c|c|c|c|c|c|c|c|c|c|} \hline
       & $\langle {\bf s}_0 \cdot{\bf s}_{\bf R}\rangle$ & $\langle {\bf s}_0 \cdot{\bf
s}_{{\bf R}}\rangle$ & $\langle {\bf s}_0 \cdot{\bf s}_{{\bf R}}\rangle$ & $\langle {\bf s}_0 \cdot{\bf
s}_{{\bf R}}\rangle$ & $\langle {\bf s}_0 \cdot{\bf s}_{{\bf R}}\rangle$  \\ \hline
 $J_2$ & ${\bf R}= (1,2)$  & ${\bf R}= (1,3)$ & ${\bf R}= (2,2)$ & ${\bf R}= (2,3)$ & ${\bf R}= (4,-2)$  \\ \hline \hline
 0.40 &  -0.027909 & -0.177408  & 0.160527  & -0.007815 &  0.154329 \\ 
 0.45 &  -0.023862 & -0.183867  & 0.176142  & -0.003522 &  0.166329 \\ 
 0.50 &  -0.021057 & -0.186297  & 0.181923  & -0.002268 &  0.169644 \\
 0.55 &  -0.018936 & -0.187719  & 0.185043  & -0.001671 &  0.171225 \\
 0.60 &  -0.017247 & -0.188748  & 0.187071  & -0.001323 &  0.172227 \\
 0.65 &  -0.015858 & -0.189567  & 0.188526  & -0.001098 &  0.172968 \\
 0.70 &  -0.014691 & -0.190248  & 0.189642  & -0.000936 &  0.173562 \\
 0.80 &  -0.012825 & -0.191325  & 0.191262  & -0.000723 &  0.174483 \\
 0.90 &  -0.011394 & -0.192135  & 0.192387  & -0.000588 &  0.175173 \\
 1.00 &  -0.010260 & -0.192765  & 0.193215  & -0.000498 &  0.175707\\
\hline
\end{tabular}
\end{center}
\end{table}
In order to analyze GS magnetic ordering, the spin-spin correlation
functions and an appropriate order parameter are important quantities.
Following Schulz et al.,\cite{schulz} we use here the 
${\bf Q}$-dependent susceptibility (square of order parameter) defined as
\be \label{M_squared} M_N^2({\bf Q})=
\frac{1}{N(N+2)}\sum_{i,j} \langle{\bf s}_i \cdot {\bf s}_j\rangle 
e^{i{\bf Q}({\bf R}_i - {\bf R}_j)}.
\ee
 The relevant order parameter for the
collinear stripe magnetic LRO present at large $J_2$ is $M_N^2({\bf Q})$
at the magnetic wave vectors ${\bf Q}_{1}=(\pi,0 )$ or ${\bf
Q}_{2}=(0,\pi )$.
In Table \ref{tab1} we give the singlet GS energies per site as well
as the order parameters $M_N^2({0,\pi})$ for finite lattices of
$N=32$, $36$, and $40$ sites. In a second Table~\ref{tab2} we present
the spin-spin correlation functions 
$\langle{\bf s}_0 \cdot {\bf s}_{\bf R}\rangle$ for the largest lattice considered. There are
in total 11 different correlation functions for $N=40$ that are
given in Table~{\ref{tab2} for the same data points as in
Table~\ref{tab1}.  Periodic boundary conditions are
imposed for all of the lattices employed here. 
A graphical representation of the finite lattices can be
found in Refs.~\cite{schulz} and \cite{ED40}.  The exact data, which is
provided here in some detail in the tables, might be used as benchmark
data for approximate methods applied to this model.\footnote{The
spin-spin correlation functions for $N=20$, $32$, and $36$ can be
provided upon request.}

The corresponding data for the energies, the square of order
parameters and for some selected correlation functions are also
displayed in more detail in Figs.~\ref{E_ED_32_40}, \ref{mag40}, and
~\ref{sisj40}.
By way of comparison, we also present corresponding curves for the AFM model
($J_1=+1$) \cite{ED40}. We recall that the GS is identical for the FM and the AFM model
on the classical level for $J_2>0.5$. For the
spin-1/2 model, due to quantum fluctuations, the two models behave
differently. Only in the limit $J_2\to \infty$ the sign of $J_1$
becomes irrelevant for the quantum GS. Our numerical data for the GS
energy, the order parameter, and the spin-spin correlation functions
demonstrate that for $J_2 \gtrsim 0.8$ the different models behave
qualitatively very similarly.  Note, however, that due the different
sign in the NN exchange $J_1$, also the NN spin-spin correlation
$\langle {\bf s}_0\cdot{\bf s}_{\bf R}\rangle$, $ {\bf R}=(1,0)$, has
a different sign for the models.  In the region $0.8 \gtrsim J_2
\gtrsim 0.6$ the differences between both models become more and more
pronounced.  For $J_2 \lesssim 0.6$ both models behave completely
different. From many previous studies
\cite{chandra88,dagotto89,figu90,schulz,richter93,richter94,Trumper97,bishop98,capriotti01,siu01,singh03,Sir:2006,Schm:2006,bishop08,darradi08,xxz,singh2009,cirac2009,ED40}
we know that $J_2 \approx 0.6$ is the point where for the AFM model
the semiclassical collinear stripe LRO gives way for a 
magnetically disordered phase.  
It is obvious that at $J_2 \approx 0.6$ the order
parameter $M_N^2({0,\pi})$ of the AFM model changes drastically (cf.\
Fig.~\ref{mag40}, whereas $M_N^2({0,\pi})$ remains almost constant up
to about $J_2 \approx 0.45$ for the FM model, i.e. there is no
indication for a breakdown of semiclassical magnetic LRO around $J_2
\approx 0.6$ for $J_1=-1$. Only for $J_2 \lesssim 0.45$ there is a
noticeable decrease in the order parameter $M_N^2({0,\pi})$ until the
transition point $J_2^{c1}$ between the singlet GS and the fully
polarized FM GS. Note, however, that $M_N^2({0,\pi})$ remains finite
until $J_2^{c1}$.  The different behavior of both models can also be
seen in the GS energy and the spin-spin correlation functions,
cf. Figs.~\ref{E_ED_32_40} and \ref{sisj40}. From
Fig.~\ref{E_ED_32_40} it is evident that the finite size-effects in
the GS energy $E_0$ of the FM model are small. In fact, the difference
in $E_0$ between $N=36$ and $N=40$ is less than $1 \%$ in the whole
region shown in Fig.~\ref{E_ED_32_40}.  Hence, a finite-size
extrapolation should give accurate results for the GS energy (see
below).

\begin{figure}  
\begin{center}
\includegraphics[clip=on,width=110mm,angle=270]{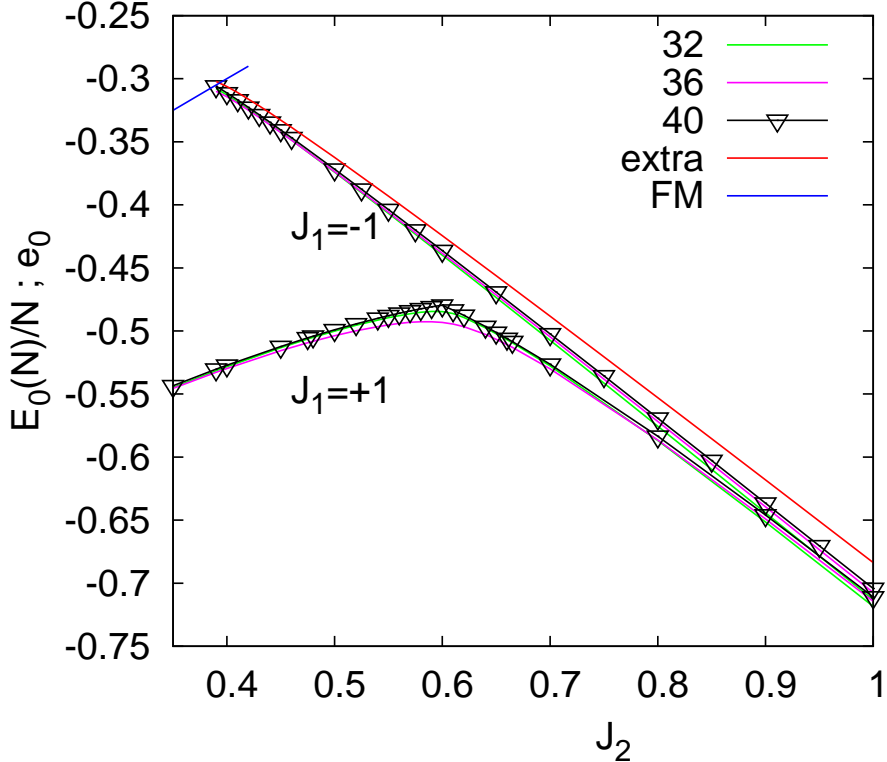}
\end{center}
\caption{GS energy per site $E_0(N)/N$ calculated by ED for finite lattices of $N=32,
36, 40$ sites as well as the extrapolated value $e_0$ ($N \to \infty$), cf. Eq.~(\ref{e}). For
comparison we show also $E_0(N)/N$ for the AFM model ($J_1=+1$) with $N=32, 36, 40$
\cite{ED40}. The blue line shows the energy of the FM eigenstate. 
}
\label{E_ED_32_40}
\end{figure}
\begin{figure} 
\begin{center}
\epsfig{file=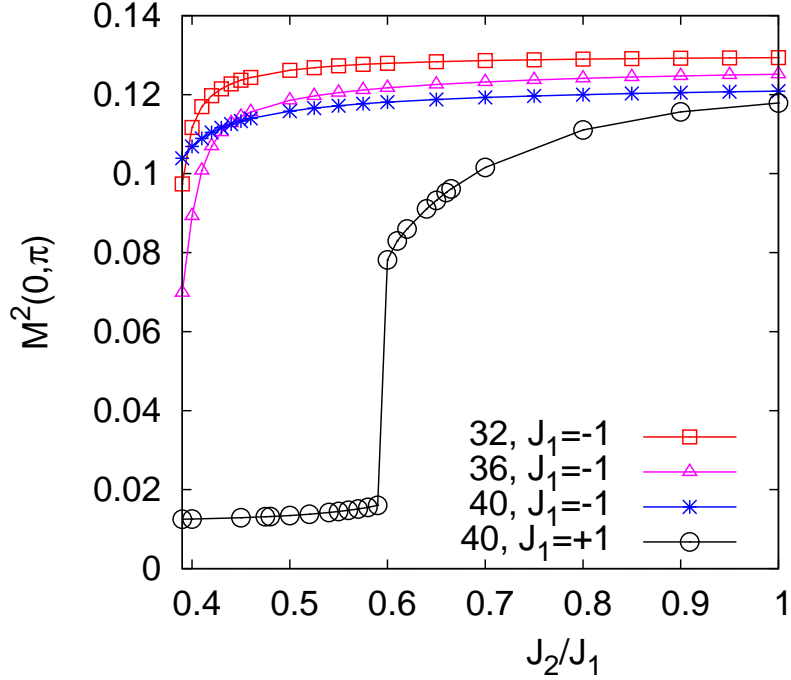,scale=1.1,angle=0.0}
\end{center}
\caption{ Square of order parameters $M_N^2(0,\pi)$, see Eq.~(\ref{M_squared}),  for
$N=32, 36,$ and $40$ ($J_1=-1$). For comparison we also present
a corresponding curve for the AFM model ($J_1=+1$) for $N=40$.
}
\label{mag40}
\end{figure}

\begin{figure} 
\begin{center}
\epsfig{file=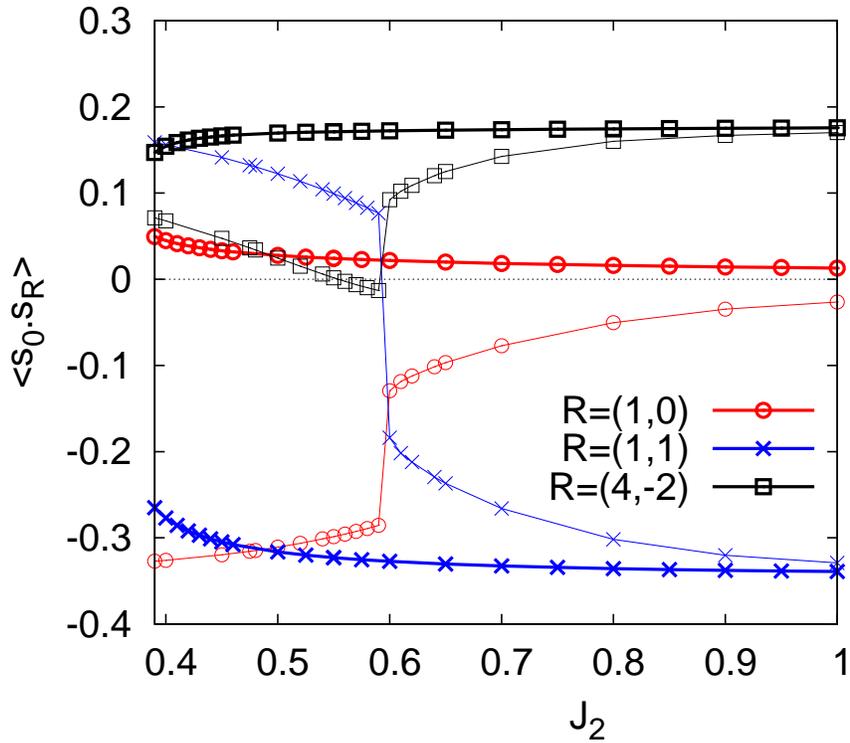,scale=0.6,angle=270.0}
\end{center}
\caption{Selected spin-spin correlation functions  $\langle {\bf s}_0
\cdot {\bf s}_{\bf R}\rangle $ calculated with Lanczos ED for $N=40$ (thick lines: $J_1=-1$, thin
lines $J_1=+1$).
}
\label{sisj40}
\end{figure}

\begin{figure} 
\begin{center}
\epsfig{file=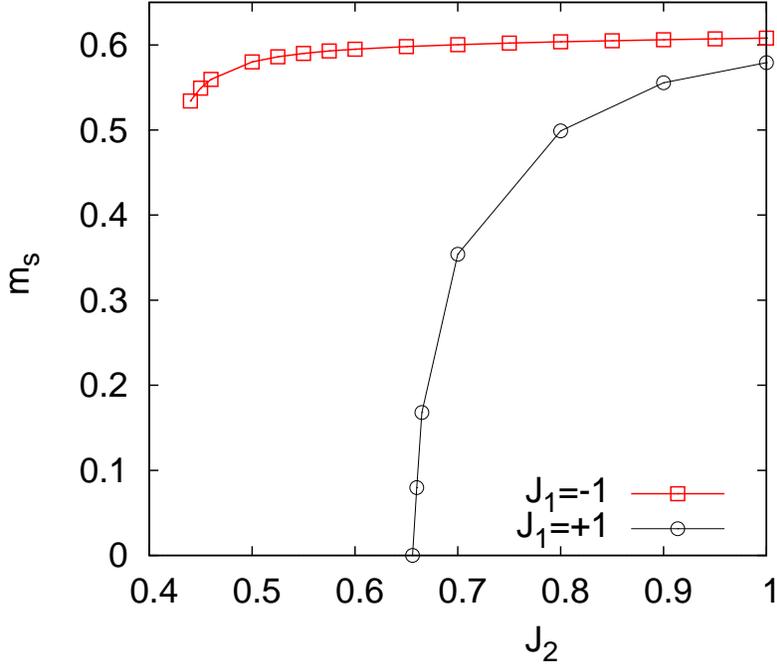,scale=1.1,angle=0.0}
\end{center}
\caption{ Extrapolated order parameters $m_s$, see Eq.~(\ref{m_s}), for the FM
model ($J_1=-1$) and the AFM model ($J_1=+1$).
}
\label{mag_extra}
\end{figure}

Following the lines of Refs.~\cite{schulz,ED40} we use now the
data for $N=20,
32, 36, 40$ to perform a finite-size extrapolation. Note that we do not include the data
for the lattice with $N=16$ sites,
since this lattice exhibits
an anomalous behavior \cite{schulz}. The finite-size extrapolation rules for 
the considered two-dimensional model are well known
\cite{neuberger,hasenfratz}. 
The finite-size behavior of  the GS energy is given by 
\be \label{e} 
\frac{E_{0}(N)}{N}= e_0 - \frac{\rm const.}{N^{3/2}} + \cdots \; .
\ee
The extrapolated energy $e_0$ is shown in Fig.~\ref{E_ED_32_40}.
We use  the intersection point between the extrapolated  singlet GS energy $e_0$ and
the energy of the fully polarized FM GS  to determine the transition point
 between
both GS phases to $J_2^{c1} = 0.393$.

The order parameter for the collinear stripe LRO in the thermodynamic
limit is defined as \cite{schulz,ED40} $m_s = \sqrt{8} \lim_{N \to
\infty} M_N(0,\pi)$.  The finite-size behavior of $M_N(0,\pi)$ is
given by \be \label{m_s} M_N^2(\pi,0)= \frac{1}{8}m_s^2 + \frac{\rm
const.}{\sqrt{N}} + \cdots \; .
\ee
However, it is evident  from Fig.~\ref{mag40} that the finite-size
behavior of $M_N^2(\pi,0)$ becomes irregular at $J_2 = 0.44$, since we
have $M_{40}^2(\pi,0) > M_{36}^2(\pi,0)$ for $J_2^{c1} < J_2 < 0.44$.
Hence, the extrapolation of $M_N^2(\pi,0)$ becomes inconclusive in
that region.  The results for $m_s$ for $J_2 \ge 0.44$ are shown in
Fig.~\ref{mag_extra}. For comparison we show also the extrapolated
order parameter for the AFM model\cite{ED40}. Although $m_s$ becomes zero at $J_2^{c2}
\approx 0.66J_1$ for the AFM model, we do not find an indication for a magnetically disordered
phase in the region above $J_2 =0.44$ for the FM model. However, we cannot
exclude such a GS phase in the very small parameter region located
between $J_2 =0.393$ and $J_2 =0.44$ from our ED data.

\begin{figure} 
\begin{center}
\includegraphics[clip=on,width=100mm,angle=270]{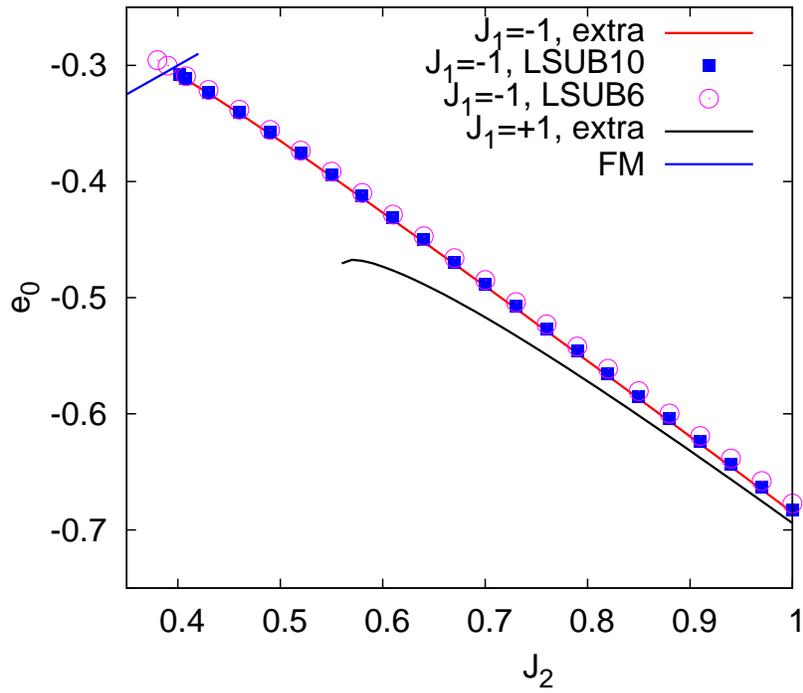}
\end{center}
\caption{The GS energy per spin as function of $J_2$
obtained by CCM-LSUB$n$ with $n=4, 6, 8, 10$
and extrapolated values in the limit $n \to \infty$ 
using the extrapolation scheme
$e(n) = a_0 + a_1(1/n)^2 + a_2(1/n)^4$.
For the sake of comparison we show also
the extrapolated GS energy for the AFM model \cite{darradi08}. The blue line
shows the energy of the FM eigenstate.
}
\label{E_CCM}
\end{figure}

\section{Results of the coupled-cluster method (CCM) for the infinite square lattice} 
\label{ccm}

For the sake of brevity, we will illustrate only some important
aspects of the CCM. The interested reader can find more details
concerning the application of the CCM on the AFM $J_1$-$J_2$ in
Refs.~\cite{bishop98,xxz,darradi08,bishop08}. For more general
information on the methodology of the CCM, see, e.g., 
Refs.~\cite{zeng98,bishop00,bishop04}.

First we notice that the CCM approach automatically yields results in 
the thermodynamic limit $N\to\infty$.  The starting point for a CCM
calculation is the choice of a normalized reference (or model) state
$|\Phi\rangle$. We then define a set of mutually commuting multispin
creation operators $C_I^+$ with respect to this state, which are 
themselves defined over a complete set of
many-body configurations $I$.  For the system under consideration here we
choose one of the two degenerate classical collinear stripe states as
the reference state.  Next, we perform a rotation of the local axis of
the spins such that all spins in the reference state align along the
negative $z$ axis.  In the rotated coordinate frame the reference state
reads $|\Phi\rangle \hspace{-3pt} = \hspace{-3pt}
|\hspace{-3pt}\downarrow\rangle |\hspace{-3pt}\downarrow\rangle
|\hspace{-3pt}\downarrow\rangle \ldots \,$, and we can treat each site
equivalently.  The corresponding multispin creation operators then can
be written as $C_I^+=s_i^+,\,\,s_i^+s_{j}^+,\,\,
s_i^+s_{j}^+s_{k}^+,\cdots$, where the indices $i,j,k,\dots$ denote
arbitrary lattice sites.

The CCM parametrizations of the ket- and bra- GS's are given
by
\begin{eqnarray}
\label{ket}
H|\Psi\rangle = E|\Psi\rangle ; 
\qquad  
\langle\tilde{\Psi}|H = E\langle\tilde{\Psi}| ;
\nonumber\\
|\Psi\rangle = e^S|\Phi\rangle, 
\qquad 
S = \sum_{I \neq 0}{\cal S}_IC_I^+ ; 
\nonumber\\
\langle\tilde{\Psi}| =  \langle\Phi|\tilde{S}e^{-S},
\qquad 
\tilde{S} = 1 + \sum_{I \neq 0}\tilde{\cal S}_IC_I^- .
\end{eqnarray}
We wish to determine the correlation coefficients ${\cal S}_I$ and $\tilde{\cal S}_I$
for the correlation operators $S$ and $\tilde {S}$. 
By using the Schr\"odinger equation, $H|\Psi\ra=E|\Psi\ra$,
we can write the GS energy as $E=\la\Phi|e^{-S}He^S|\Phi\ra$.  The
magnetic order parameter is given by
\be 
\label{mag}
m_s = -\frac{1}{Ns} \sum_{i=1}^N \la\tilde\Psi|s_i^z|\Psi\ra ,
\ee
where $s_i^z$ is expressed in the rotated coordinate system and
$s=1/2$ is the spin quantum number. To find the ket-state and
bra-state correlation coefficients we have to solve the so-called CCM
ket- and bra-state equations given by
\begin{eqnarray}
\label{ket_eq}
\langle\Phi|C_I^-e^{-S}He^S|\Phi\rangle = 0,
\qquad 
\forall I\neq 0,
\\
\label{bra_eq}
\langle\Phi|\tilde{\cal S}e^{-S}[H, C_I^+]e^S|\Phi\rangle = 0,
\qquad 
\forall I\neq 0.
\end{eqnarray}
Each ket- or bra-state equation belongs to a certain creation operator
$C_I^+=s_i^+,\,\,s_i^+s_{j}^+,\,\, s_i^+s_{j}^+s_{k}^+,\cdots$,
i.e. it corresponds to a certain set (configuration) of lattice sites
$i,j,k,\dots\;$.

For the considered quantum many-body model we have to use approximations
in order to truncate the expansion of $S$ and $\tilde {S}$.
We use the well established LSUB$n$ scheme
\cite{zeng98,krueger00,bishop00,bishop04,rachid05,bishop08,xxz,darradi08}
in which all multispin correlations in the correlation operators $S$ and 
$\tilde {S}$ over all distinct locales on the lattice
defined by $n$ or fewer contiguous sites are taken into account.  For
instance, within the LSUB4 approximation, multispin creation operators
of one, two, three or four spins distributed on arbitrary clusters of
four contiguous lattice sites are included.  The number of these fundamental 
configurations can be reduced by using lattice symmetry
and conservation laws.  In the highest order of approximation considered
here, namely the LSUB10 approximation, we have finally $45825$
fundamental configurations for the collinear stripe reference state,
yielding $45825$ coupled nonlinear equations which have to be solved
numerically.

The LSUB$n$ approximation becomes exact for $n \to \infty$, and so we
can improve our results by extrapolating the ``raw'' LSUB$n$ data to
$n \to \infty$.  There is ample experience regarding how one should 
extrapolate the GS energy $e_0(n)$ and the magnetic order parameter 
$m_s(n)$. For the GS energy per spin $e_0(n) = a_0 + a_1(1/n)^2 + a_2(1/n)^4$ is a
reasonable well-tested extrapolation ansatz
\cite{krueger00,bishop00,bishop04,rachid05,
Schm:2006,bishop08,xxz,darradi08}.  An appropriate extrapolation rule
for the magnetic order parameter is
$m_s(n)=b_0+b_1(1/n)^{1/2}+b_2(1/n)^{3/2}$ \cite{bishop08,xxz,darradi08}.
Moreover, we know from Refs.~\cite{bishop08,xxz,darradi08} that the lowest level
of approximation called the LSUB2 approximation conforms poorly to these
rules. Hence, as in previous calculations \cite{bishop08,xxz,darradi08}, we
exclude LSUB2 data from the extrapolations.

\begin{figure} 
\begin{center}
\epsfig{file=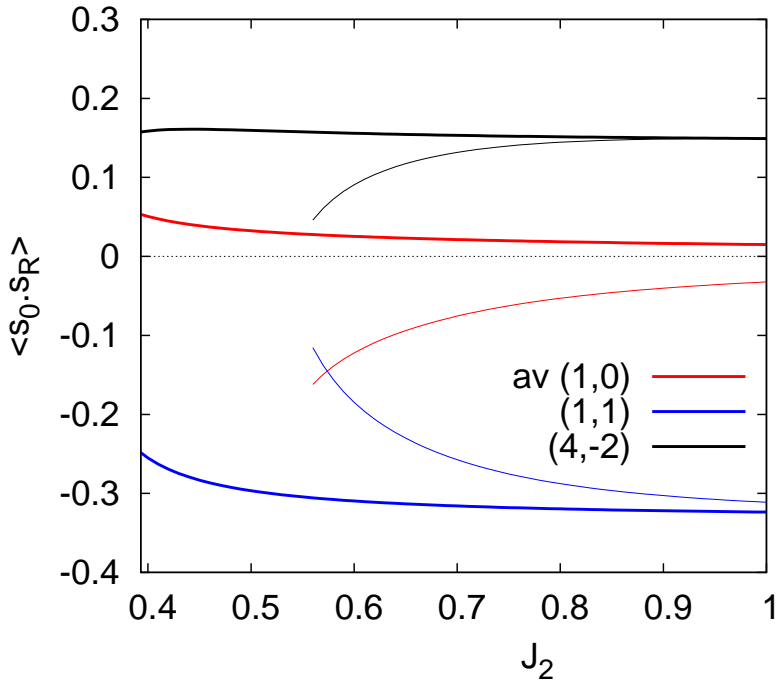,scale=1.1,angle=0.0}
\end{center}
\caption{Selected spin-spin correlation functions  
$\langle {\bf s}_0 \cdot {\bf s}_{\bf R}\rangle $ calculated  with CCM within LSUB6
approximation (thick lines: $J_1=-1$, thin
lines $J_1=+1$).
}
\label{sisjccm}
\end{figure}

\begin{figure} 
\begin{center}
\epsfig{file=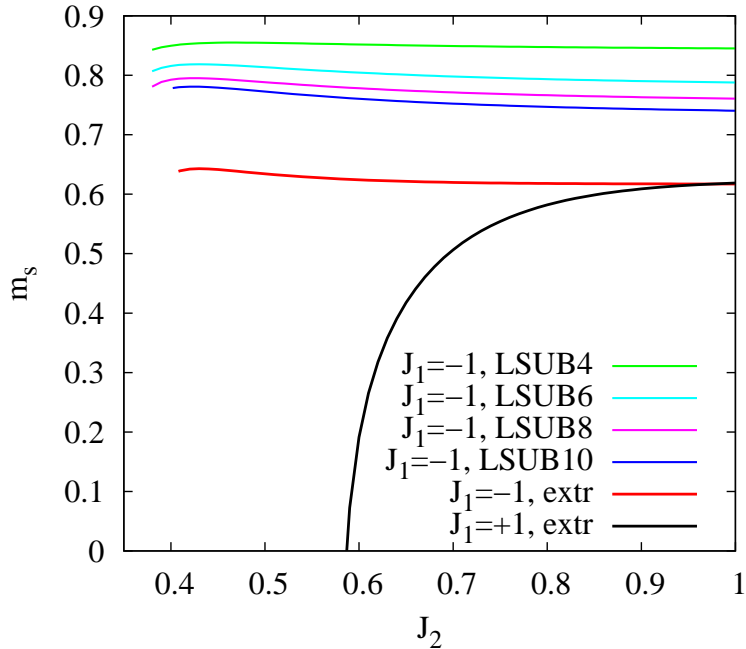,scale=1.05,angle=0.0}
\end{center}
\caption{Magnetic order parameter $m_s$ versus $J_2$ obtained by CCM-LSUB$n$
with $n=4, 6, 8, 10$ and its extrapolated values ($n \to \infty$)
using the extrapolation scheme
$m_s(n)=b_0+b_1(1/n)^{1/2}+b_2(1/n)^{3/2}$. By way of comparison we show
also the extrapolated collinear stripe order parameter ($n \to
\infty$) for the AFM model ($J_1=+1$).}
\label{m_CCM}
\end{figure}

Note that starting from large $J_2$ the solution of the CCM-LSUB10
equations can be traced until $J_2=0.402$, only. 
For $J_2 < 0.402$ no
CCM-LSUB10 solutions with respect to the collinear stripe reference state
could be found.   This could be related to
the complexity of the set of 45825 coupled equations and the fact  that
the fully polarized FM state is a low-lying excitation very close to the true 
GS for $J_2 \sim 0.4$.
However, we cannot exclude the possibility 
that this  problem  might also be an indication that the
collinear stripe order is not a good description for the true GS 
for $J_2$ values very close to the transition to the FM GS.

In Fig.~\ref{E_CCM} we show the GS energy per site for the LSUB6 and
LSUB10 approximations, as well as the extrapolated GS energy ($n \to \infty$). The
numerical CCM-LSUB$n$ data demonstrate, that the difference between
the various LSUB$n$ data is very small (see Fig.~\ref{E_CCM}).  As a
result, the extrapolated GS energy  almost coincides
with LSUB10 data (in fact, the difference is less than $0.4
\%$). Moreover, the difference between the extrapolated CCM energy and
the finite-size extrapolated ED energy is less than $1\%$ in the whole
parameter region. Hence, we conclude that the GS energy data presented
in this paper for the FM $J_1$-$J_2$ model are very accurate.  
We again use the intersection point of the energies in order to find
an estimate for the transition point $J_2^{c1}$ to the FM GS.
We find that $J_2^{c1}=0.395$ for the LSUB6 approximation and this
is shown in Fig.~\ref{E_CCM}. We find also that $J_2^{c1}=0.394$ and 
$J_2^{c1}=0.398$ for the LSUB8 and LSUB4 approximations,
respectively (not shown in Fig.~\ref{E_CCM}). 
As the difference between the LUSB6 and LSUB8 results  for $J_2^{c1}$ is
already very small we may choose the LSUB8 value, $J_2^{c1}=0.394$, as the CCM 
estimate for the transition point. 
This CCM critical value is in excellent agreement with the ED
estimate $J_2^{c1}=0.393$ (see above).  We also compare the GS
energies for the FM and the AFM models in Fig.~\ref{E_CCM}. As for the ED
method, the deviation in energies between the two models (i.e., negative 
and positive $J_1$) becomes substantial at around $J_2=0.6$.
 
We also present the CCM results for spin-spin correlation functions for the
FM as well as the AFM model in
Fig.~\ref{sisjccm}, where we have shown the
same correlation functions as in Fig.~\ref{sisj40}. Recall that one of the two
degenerate collinear stripe states has to be chosen  as CCM reference
state. As a result the NN
correlation functions $\langle {\bf s}_0 \cdot {\bf s}_{\bf R}\rangle $, ${\bf
R}=(1,0)$, and 
$\langle {\bf s}_0 \cdot {\bf s}_{\bf R}\rangle $, ${\bf R}=(0,1)$, are
different, whereas, e.g., all NNN correlation functions are the same. 
Hence, we show in Fig.~\ref{sisjccm} the averaged NN correlation
function.
For the FM model the agreement of the CCM correlation functions  with the ED data is
excellent, see Figs.~\ref{sisj40} and \ref{sisjccm}. For the AFM model there
is a noticable difference for  $J_2<0.59$. In particular, the steep decrease in 
the correlation functions at $J_2=0.59$ 
present in the ED data for the AFM model, cf. Ref.~\cite{ED40}, is not observed 
in the CCM LSUB6 data. For $J_2<0.59$ most likely no magnetic LRO
exists for the AFM model\cite{darradi08}. Therefore it is not surprising
that  the CCM 
solution in a finite order of LSUB$n$ approximation based on the collinear stripe
reference state  does not povide quantitatively precise results for the correlation
functions inside the magnetically disordered phase.

Last but not least, we present results for the collinear stripe
order parameter $m_s$ calculated by the CCM via Eq.~(\ref{mag}) in Fig.~\ref{m_CCM}. 
For the order parameter, the CCM-LSUB$n$ results depend on the
approximation level $n$ more strongly than for the GS energy. Hence, it
is more important to extrapolate them in the limit $n \to \infty$.  
By comparing the correlation functions and the order parameter of the FM model
with the AFM model, it is again obvious that both models behave
similarly for larger $J_2$. However, there are significant
differences for $J_2 \lesssim 0.6 $ between both models. Although the
CCM result for the AFM model clearly shows the breakdown of the
semiclassical collinear magnetic LRO around $J_2=0.6$ (the CCM
estimate for the critical frustration is $J_2^{c2}\approx
0.59$\cite{darradi08}), the spin-spin correlation functions and the  
order parameter for the FM model stay almost constant over the 
whole parameter region shown in Fig.~\ref{m_CCM}. 
Again, we obtain good agreement of the order parameter between the ED data
and the CCM data in a wide parameter range, cf. Figs.~\ref{mag_extra}
and \ref{m_CCM}. 
Finally, we
come to a similar conclusion as in Sec.~\ref{ED} that we do not find
an indication for the breakdown of the collinear stripe magnetic LRO
until $J_2 \sim 0.4$ for the FM model.

\section{Summary}

In this paper we have presented results on the GS magnetic ordering 
of the $J_1$-$J_2$ spin-$1/2$
Heisenberg magnet with FM NN exchange $J_1=-1$ on the square lattice obtained by Lanczos
ED as well as the CCM.
In agreement with previous
studies\cite{shannon04,shannon06,sindz07,sousa,haertel10,dmitriev97} we obtain 
values for the transition point $J_2^{c1}$ at which the     
fully polarized FM GS (present for small $J_2$) breaks down,
$J_2^{c1} = 0.393|J_1|$ (ED) and $J_2^{c1} = 0.394|J_1|$ (CCM).

In contrast to previous
studies\cite{shannon04,shannon06,sindz07,sousa} we do not find
indications for the breakdown of the semiclassical collinear magnetic
LRO (present for large $J_2$) at about $J_2=0.6|J_1|$, rather this
magnetic LRO seems to be stable till $J_2 \sim 0.4|J_1|$.

Although our results are in favour of semiclassical magnetic GS long-range order
for a wide range of frustrating exchange $J_2$, the low-lying
excitations and, as a consequence, the low-temperature thermodynamics
might be strongly influenced by frustration. This becomes increasingly important 
when approaching the transition to the FM GS at $J_2^{c1}\approx
0.4\vert J_1 \vert$, see e.g. Ref.~\cite{haertel10}. For $J_2
\gtrsim J_2^{c1}$ the FM multiplet becomes a low-lying
excitation, and so an additional low-energy scale will appear leading, e.g.,
to an extra low-temperature peak in the specific heat, see, e.g.,
Refs. \cite{shannon04,heidrich06,tmrg}. 

Stimulated by the recent experimental activities and the related search
for new square-lattice materials near the quantum critical point of the
$J_1$-$J_2$ model, the issue of low-lying excitations and their
relevance for the thermodynamic properties should be investigated in
future studies in more detail.\\

{\it \bf  Acknowledgement:}
We thank S.E. Kr\"uger and R. Zinke for valuable discussions. 
This work was supported by the DFG (Ri615/16-1).



\begin{thebibliography}{99}
\bibitem{chandra88} P.~Chandra and B.~Doucot,  { Phys. Rev. B} {\bf 38},
  9335 (1988).

\bibitem{dagotto89} E.~Dagotto and A.~Moreo,  { Phys. Rev. Lett.} {\bf 63},
  2148 (1989).
\bibitem{figu90}
F.~Figueirido, A. Karlhede, S. Kivelson, S. Sondhi, M.~Rocek, and
D.S.~Rokhsar,
Phys. Rev. B {\bf 41}, 4619 (1990).

\bibitem{schulz} H.J.~Schulz and T.A.L.~Ziman, { Europhys. Lett.}
  {\bf 18}, 355 (1992);
H.J. Schulz, T.A.L.~Ziman, and D.~Poilblanc, { J.~Phys.~I}
  {\bf 6}, 675 (1996).
\bibitem{richter93}
   J.~Richter, Phys. Rev. B {\bf 47}, 5794 (1993).
\bibitem{richter94} 
J.~Richter, N.B.~Ivanov, and K.~Retzlaff,
      Europhys. Lett. {\bf 25}, 545 (1994).
\bibitem{Trumper97}
A.~E.~Trumper, L.~O.~Manuel, C.~J.~Gazza, and H.~A.~Ceccatto,
Phys. Rev. Lett. {\bf 78}, 2216 (1997).
\bibitem{bishop98} R.F.~Bishop, D.J.J.~Farnell, and J.B.~Parkinson,
 {  Phys. Rev. B} {\bf 58}, 6394 (1998).
\bibitem{capriotti01} L.~Capriotti, F.~Becca, A.~Parola, and S.~Sorella, 
{ Phys. Rev. Lett.} {\bf 87}, 097201 (2001).
\bibitem{siu01} L. Siurakshina, D. Ihle, and R. Hayn, Phys. Rev. B {\bf 64},
104406 (2001).
\bibitem{singh03} R.R.P.~Singh, Weihong Zheng , J.~Oitmaa, 
O.P.~Sushkov, and C.J.~Hamer,
{ Phys. Rev. Lett.} {\bf 91}, 017201 (2003).
\bibitem{Sir:2006} J. Sirker, Z. Weihong, O. P. Sushkov, and J. Oitmaa,
Phys. Rev. B {\bf 73}, 184420 (2006).
\bibitem{Schm:2006} D. Schmalfu{\ss}, R. Darradi, J. Richter,
J.~Schulenburg, and D.~Ihle,
Phys. Rev. Lett. {\bf 97}, 157201 (2006).
\bibitem{bishop08} R.F.~Bishop, P.H.Y.~Li, R.~Darradi, and J. Richter,
      J. Phys.: Condens. Matter {\bf 20}, 255251 (2008).

\bibitem{xxz} R.F.~Bishop, P.H.Y.~Li, R.~Darradi, J.~Schulenburg and J.
Richter,
     Phys. Rev. B {\bf 78}, 054412 (2008).                    

 \bibitem{darradi08} R.~Darradi, O.~Derzhko, R.~Zinke, J.~Schulenburg,  S.~E.~Kr\"uger, and
       J.~Richter,
       Phys. Rev. B {\bf 78}, 214415 (2008).

\bibitem{singh2009} T. Pardini and R.R.P. Singh,
Phys. Rev. B {\bf 79}, 094413 (2009).

\bibitem{cirac2009} 
V. Murg, F. Verstraete, and J. I. Cirac, Phys. Rev. B {\bf 79},
195119 (2009).

\bibitem{ED40}J. Richter and J. Schulenburg,
Eur. Phys. J. B {\bf 73}, 117 (2010).

\bibitem{fprg} J. Reuther and P. Woelfle, arXiv:0912.0860.

\bibitem{kaul04} E.~E.~Kaul, H.~Rosner, N.~Shannon, R.V.~Shpanchenko, and C.~ Geibel, 
J. Magn. Magn. Mater. \textbf{272-276(II)}, 922 (2004).
\bibitem{jmmm07} M. Skoulatos, J.P. Goff, N. Shannon, E.E. Kaul, C. Geibel, A.P. Murani, M.
Enderle, and  A.R. Wildes,
J. Magn. Magn. Mater. \textbf{310}, 1257 (2007).
\bibitem{carretta2009}
P. Carretta,  M. Filibian,  R. Nath,  C. Geibel, and P. J. C. King,
Phys. Rev. B {\bf 79}, 224432 (2009).
\bibitem{enderle}
M. Skoulatos, J.P. Goff, C. Geibel, E.E. Kaul, R. Nath, N. Shannon,
B. Schmidt, A.P. Murani, P.P. Deen, M. Enderle, and A.R. Wildes,
arXiv:0909.0702.

\bibitem{kageyama05} H.~Kageyama, T.~Kitano, N.~Oba, M.~Nishi, S.~Nagai, K.~Hirota, L.~Viciu, J.B.~Wiley, J.~Yasuda, Y.~Baba, Y.~Ajiro, 
and K.~Yoshimura, J. Phys. Soc. Jpn. \textbf{74}, 1702 (2005).

\bibitem{rosner09}
A.A. Tsirlin and H. Rosner,
Phys. Rev. B {\bf 79}  214417  (2009).
\bibitem{rosner09a}
 A.A. Tsirlin, B. Schmidt,  Y. Skourski,
   R. Nath, C. Geibel, and H. Rosner,
Phys. Rev. B {\bf 80} 132407  (2009).
\bibitem{nath2008}
R. Nath, A.A. Tsirlin, H. Rosner, and C. Geibel,
Phys. Rev. B {\bf 78}  064422 (2008).


\bibitem{shannon04} N.~Shannon, B.~Schmidt, K.~Penc, and P.~Thalmeier, Eur. Phys. J. B \textbf{38}, 599
(2004).
\bibitem{shannon06} N.~Shannon, T.~Momoi, and P.~Sindzingre, Phys. Rev. Lett. \textbf{96}, 027213
(2006).
\bibitem{sindz07}
P. Sindzingre, N. Shannon and T. Momoi, J. Magn. Magn. Mat.
{\bf  310}, 1340 (2007).
\bibitem{schmidt07} B.~Schmidt, N.~Shannon, and P.~Thalmeier, J. Phys. Cond. Mat. \textbf{19}, 145211 (2007).
\bibitem{schmidt07_2} B.~Schmidt, N.~Shannon, and P.~Thalmeier, J. Magn. Magn. Mater. \textbf{310}, 1231
(2007).
\bibitem{sousa}  J.R.~Viana and J.R.~de~Sousa,
 Phys.\ Rev.\ B {\bf 75}, 052403 (2007). 
\bibitem{shannon09} P. Sindzingre, L. Seabra, N. Shannon, and
T. Momoi,
J. Phys.: Conf.  Series {\bf 145}, 012048 (2009).
\bibitem{sindz09}  P. Sindzingre, N. Shannon, and
T. Momoi, arXiv:0907.4163.
\bibitem{haertel10}
M. H\"artel, J. Richter, D.~Ihle, and S.-L.~Drechsler,
arXiv:1001.1222.
 
\bibitem{dmitriev97}
D.~V.~Dmitriev, V.~Ya.~Krivnov, and A.~A.~Ovchinnikov, Phys. Rev. B
\textbf{55}, 3620 (1997). 

\bibitem{Mune:2007}
T. Munehisa and Y. Munehisa,
J. Phys: Condensed Matter {\bf 19}, 196202 (2007).
 
\bibitem{Kash:2001}
T. Kashima and M.Imada,
J. Phys. Soc. Jpn. {\bf 70}, 3052 (2001). 


\bibitem{wir04} J.~Richter,  J.~Schulenburg, and A.~Honecker,
      in: \textit{Quantum Magnetism},
      ed by U.~Schollw\"ock, J.~Richter, D.J.J.~Farnell, R.F.~Bishop, 
      Lecture Notes in Physics {\bf 645}, p. 85 (Springer, Berlin, 2004).


\bibitem{Lauch:2004}
A. L\"auchli, J.C. Domenge, C. Lhuillier, P. Sindzingre, and M.~Troyer,
Phys. Rev. Lett. {\bf 95}, 137206 (2005).

\bibitem{star04} J. Richter, J. Schulenburg, A. Honecker, and
D.~Schmalfu{\ss},
      Phys. Rev. B {\bf 70}, 174454 (2004).


\bibitem{prl02}
J.~Schulenburg, A.~Honecker, J.~Schnack, J.~Richter, 
and H.~-~J.~Schmidt,
Phys. Rev. Lett. {\bf 88,} 167207 (2002); 
A. Honecker,  J. Schulenburg, and J.Richter,
            J. Phys.: Condens. Matter {\bf 16}, S749 (2004).


\bibitem{neuberger}
H.~Neuberger and T.~Ziman,
Phys. Rev. B {\bf 39}, 2608 (1989).

\bibitem{hasenfratz}
   P. Hasenfratz and F. Niedermayer, Z. Phys. B: Condens. Matter
    {\bf 92}, 91 (1993).


\bibitem{zeng98}
C.~Zeng, D.~J.~J.~Farnell, and R.~F.~Bishop,
J. Stat. Phys. {\bf 90}, 327 (1998).

\bibitem{bishop00} 
R.~F.~Bishop, D.~J.~J.~Farnell, S.~E.~Kr\"uger,  J.~B.~Parkinson, and
J.~Richter,
J. Phys.: Condens. Matter { \bf 12}, 6877 (2000).
\bibitem{bishop04}
D.~J.~J.~Farnell and R.~F.~Bishop,
in {\it Quantum Magnetism}, Lecture Notes in Physics Vol. {\bf 645}, 
edited by U.~Schollw\"ock, J.~Richter, D.~J.~J.~Farnell, and R.~F.~Bishop (Springer, Berlin, 2004), p. 307. 


\bibitem{krueger00} 
S.~E.~Kr\"uger,  J.~Richter, J.~Schulenburg, D.~J.~J.~Farnell, and
R.~F.~Bishop,
Phys. Rev. B {\bf 61}, 14607 (2000).

\bibitem{rachid05} R. Darradi, J. Richter, and D.J.J.~Farnell,
      Phys. Rev. B {\bf 72}, 104425 (2005).
\bibitem{heidrich06} F. Heidrich-Meisner, A. Honecker, and T. Vekua,
Phys. Rev. B \textbf{74}, 020403(R) (2006).
\bibitem{tmrg} H. T. Lu, Y. J. Wang, Shaojin Qin, and T. Xiang,
Phys. Rev. B \textbf{74}, 134425 (2006).


\end{thebibliography}
\end{document}